\documentstyle[epsfig,multicol,prb,aps]{revtex}
\begin{document}
\title{Calculated thermoelectric properties of La-filled skutterudites}
\author{D.J. Singh$^1$ and  I.I. Mazin$^{1,2}$}
\address{$^1$Code 6691, Naval Research Laboratory, Washington, DC 20375\\
$^2$CSI, George Mason University}
\date{May 7, 1997}
\maketitle

\begin{abstract}
The thermoelectric properties of La-filled skutterudites are discussed from
the point of view of their electronic structures. These are calculated from
first principles within the local density approximation. The electronic
structure is in turn used to determine transport related quantities. Virtual
crystal calculations for La(Fe,Co)$_4$Sb$_{12}$ show that the system obeys
near rigid band behavior with varying Co concentration, and has a
substantial band gap at a position corresponding to the composition LaFe$_3$%
CoSb$_{12}$. The valence band maximum occurs at the $\Gamma $ point and is
due to a singly degenerate dispersive  band, which by itself would
not be favorable for high thermopower.
However, very flat transition metal derived bands
occur in close proximity and become active as the doping level is increased,
giving a non-trivial dependence of the properties on carrier concentration
and explaining the favorable thermoelectric properties.
\end{abstract}

\begin{multicols}{2}There has been a recent revival of activity in the
search for improved thermoelectric (TE) materials, with an emphasis on novel
materials concepts \cite{1}. The utility of a material for TE applications
is measured by its dimensionless figure of merit, $ZT=(\sigma /\kappa )S^2T$%
, where $T$ is temperature, $\sigma $ is the electrical conductivity, $S$ is
the Seebeck coefficient and $\kappa $ is the thermal conductivity, which
contains both electronic and lattice contributions, $\kappa _e$ and $\kappa
_l$, respectively. High values of $S$ are characteristic of highly
doped semiconductors;
unfortunately, these have low electric conductivity and thus small ratio $%
(\sigma /\kappa ).$ On the other hand, metals usually have this ratio close
to the Wiedemann-Franz value, $%
\sigma /\kappa _e.$ If the Wiedemann-Franz law holds, $S=160$ V/K is required
for the state-of-the-art value $ZT=1$. Much TE research over the past 40
years has focussed on covalent semiconducting compounds and alloys, composed
of 4th and 5th row elements, with a view to finding low thermal conductivity
materials that have reasonable carrier mobilities and high band masses. Most
current generation TE materials are of this type, e.g. Bi$_2$Te$_3$, Si-Ge,
PbTe. Despite research efforts spanning three decades, little progress in
increasing $ZT$ has been achieved, and in particular Bi$_2$Te$_3$/Sb$_2$Te$_3
$ has remained the material of choice for room temperature applications.

Recently, three new materials with $ZT\geq 1$ have been reported \cite{2,4,3}%
, and these do not clearly fall in the same class as previous TE materials. $%
\beta $-Zn$_4$Sb$_3$, with reported $ZT$ up to 1.3
has a large region of linear temperature dependence
of the resistivity, suggestive of a metallic rather than semiconducting
material, but unlike normal metals this is accompanied by high thermopowers.
CeFe$_4$Sb$_{12}$ both by itself and doped with Co shows high values of $ZT$
combined with very low thermal conductivities, and depending on the
composition can show either metallic-like or semiconducting temperature
dependencies of the resistivity. At low temperature, CeFe$_4$Sb$_{12}$
displays properties reminiscent of heavy fermion materials\cite{5}.
Band structure calculations \cite{6}  show that the Ce $f$%
-states, indeed,  contribute significantly to the electronic structure near
the Fermi energy, leading the enhanced band masses, favorable for TE, and
producing band gaps via hybridization with the valence states. This would
imply a rather different electronic structure for LaFe$_4$Sb$_{12}$.
However, La(Fe,Co)$_4$Sb$_{12}$ does have $ZT\approx 1$ for appropriate
conditions. 
 Here we report first
principles calculations for La(Fe,Co)$_4$Sb$_{12}$ within the local density
approximation (LDA), similar to our previous calculations for binary and
Ce-filled skutterudites \cite{6,7,8}. We then compute transport
properties, based on the calculated band structure, and show them to be in
accord with the experiments. An analysis of the underlying band structure
reveals the mechanism for high thermopower, and suggests ways to optimize
the TE properties.

The calculations were performed using the general potential linearized
augmented planewave (LAPW) method \cite{9}. This method makes no shape
approximations to either the potential or charge density and uses flexible
basis sets in all regions of space. As such it is well suited to materials
with open crystal structures and low site symmetries like La(Fe,Co)$_4$Sb$%
_{12}$. Well converged basis sets of  $\approx$2100
functions were used. Local
orbital extensions to the basis set were used
 in order to relax linearization errors
generally and to include the upper core states of La consistently with the
valence states.

The calculations were based on the experimental crystal structure of LaFe$_4
$Sb$_{12}.$ The electronic structures of La(Fe,Co)$_4$Sb$_{12}$ alloys were
calculated as for LaFe$_4$Sb$_{12}$ using the virtual crystal method, i.e.
self-consistent LAPW calculations were performed using average ions with
fractional charges intermediate between Fe and Co. This approximation
includes the average ionic charge and band filling, but is known to distort
the electronic structure for alloys when the scattering properties of the
ions at the mixed site differ substantially. Calculations were performed for
0\%, 25\% and 50\% Co substitution. The calculated bands near the Fermi
level differ little, providing {\it aposteriori} justification of the
virtual crystal approximation.
 Transport properties relevant to TE were determined from
the calculated band structures using standard kinetic theory as given by
Ziman and others\cite{10,11}, 
\begin{eqnarray}
\sigma (T) &=&\frac{e^2}3\int d\epsilon N(\epsilon )v^2(\epsilon )\tau
(\epsilon ,T)\left( -\frac{\partial f(\epsilon )}{\partial \epsilon }\right) 
\nonumber\\
S(T) &=&\frac e{3T\sigma (T)}\int d\epsilon N(\epsilon )v^2(\epsilon
)\epsilon \tau (\epsilon ,T)\left( -\frac{\partial f(\epsilon )}{\partial
\epsilon }\right)
\label{S}\\
& =&\frac 1{3eT\sigma (T)}\int d\epsilon \sigma (\epsilon
,T)\epsilon \left( -\frac{\partial f(\epsilon )}{\partial \epsilon }\right) .
\nonumber
\end{eqnarray}
Here $N(\epsilon )$ is the density of electronic states at the energy $%
\epsilon $ per unit volume, $\tau $ is the scattering rate for electrons,
and the average velocity $v(\epsilon )$ of electrons with the energy $%
\epsilon $ is defined via 
\begin{eqnarray}
N(\epsilon )v^2(\epsilon ) &=&(2\pi )^{-3}\int \delta (\epsilon _{{\bf k}%
}-\epsilon )d{\bf k}  \label{N} \\
N(\epsilon )v^2(\epsilon ) &=&(2\pi )^{-3}\int v_{{\bf k}}^2\delta (\epsilon
_{{\bf k}}-\epsilon )d{\bf k}  \nonumber
\end{eqnarray}

The above formulae are readily calculated from the band
structure, provided that the scattering rate $\tau$ is known.  Fortunately,
in most cases $\tau$ does not vary too much with the energy, although
 there are
exceptions to this rule, e.g. Pd metal\cite{13} where $E_F$ occurs near a
very sharp feature in $N(\epsilon )$ and Kondo systems where there is
resonant scattering\cite{11}. La(Fe,Co)$_4$Sb$_{12},$ on the other hand,
shows no indications of such behavior, and $\tau (\epsilon )$ can be
replaced by a constant which then cancels out in many transport
properties, like Seebeck coefficient or Hall number.

\begin{figure}[tbp]
\centerline{\epsfig{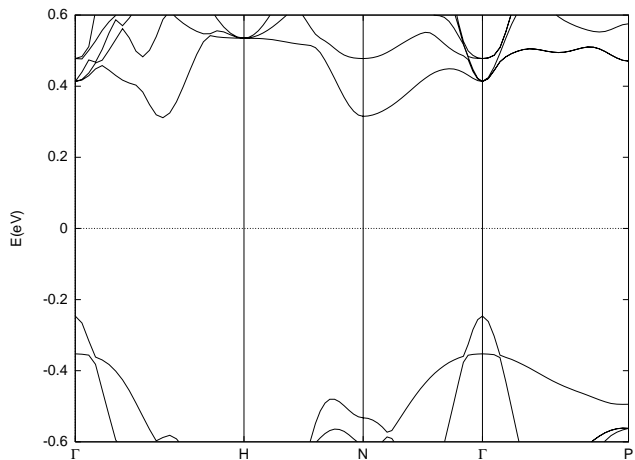}}
\vspace{0.1in} \setlength{\columnwidth}{3.2in} \nopagebreak
\caption{Virtual crystal
band structure of La(Fe$_{0.75}$Co$_{0.25})_3$Sb$_{12}$ in the vicinity of 
the Fermi level.}
\label{bands} \end{figure}

The calculated band structure of
La(Fe$_{0.75}$Co$_{0.25})_3$Sb$_{12}$ near the Fermi level is
shown in Fig. \ref{bands}. Although the initial idea of alloying unfilled
skutterudites with La\cite{1} was to improve the figure of merit by
reducing the lattice thermal conductivity, it turns out that La plays a
positive role from the point of view of the electronic structure as well.
First principles calculations for CoSb$_3$ \cite{7} show a small band gap
with a highly non-parabolic, quasi-linear valence band dispersion. This
result, which now has experimental support \cite{18,19,20}, combined with
the kinetic transport theory above, implies that in p-type CoSb$_3$ the
doping level, $n$, dependence of $S$ and $\sigma $ will differ from that of
a semiconductor with parabolic bands. In particular, in the constant
scattering time, degenerate regime, $S\propto n^{-1/3}$ and $\sigma
\propto n^{2/3}$ instead of the usual $S\propto n^{2/3}$ and $\sigma
\propto n$. This means that the power factor, $S^2$ will be less
dependent on doping level than in a conventional case and that it will be
more difficult to optimize $ZT$ by adjusting $n$ (but note that $\kappa $
has an electronic component proportional to $\sigma $).

 The calculated band
structure
of La(Fe$_{0.75}$Co$_{0.25})_3$Sb$_{12}$
  shows an indirect ${\bf \Gamma -N}$ gap of 0.60.  This is
qualitatively different from the band structure of CoSb$_3$. The direct gap
at $\Gamma$
 is 0.76 eV. Both the conduction and valence band edges are formed from
parabolic bands with mostly Sb $p-$character. In addition there are some
rather flat (heavy mass) primarily Fe/Co $d-$derived bands near but not at
the band edge both above and below the gap.

To understand this band structure we first look at the crystal structure. A
useful starting point is a hypothetical LaFe compound with the CsCl (B2)
structure. We then construct 2$\times 2\times 2$ supercell, keep two La
atoms along (111) directions, and remove the remaining six La's. The
resulting pores are filled with the Sb$_4$ rings, oriented in such a way as
to keep the (111) threefold axis. The Sb$_4$ rings form quasimolecules,
since the Sb-Sb bonds are the shortest in the crystal ($\approx 2/3$ of the
Fe-Fe bond length). 12 $p-$states of Sb atoms in a molecule hybridize with
each other forming four strongly bound $pp\sigma $ bonding states (split
into two close pairs by weak $pp\pi $ interaction), four $pp\sigma $
antibonding states, and two pairs of weakly bound $pp\pi $
bonding/antibonding states, formed by the orbitals perpendicular to the
plane of the ring. Half of these states are occupied by 6 $p-$electrons of
 four Sb atoms.
 Fe/Co $d-$states are close in energy to the Sb $p$-states. They give
rise to two sets of bands derived from 6 $t_{2g}$ states at lower energy and
4 $e_g$ states at higher energy per metal ion. Three Fe and one Co together
have 33 $d-$electrons, of which 24 reside in the $t_{2g}$ bands and 9,
together with three La electrons, fill up twelve more states in Sb rings.
Thus, the highest occupied states in the 
ionic picture occur between antibonding $%
pp\pi $ Sb bands and $t_{2g}$ Fe/Co states. Let us first discuss the former.
The symmetry of the relevant combination of $p-$orbitals with respect to the
center of the ring is $xyz,$ that is, an $f-$symmetry. Not surprisingly,
these combinations hybridize with the La $xyz$ states, forming a quasi-$%
ff\delta $ band with the maximum at the ${\bf \Gamma }$ point, where the $%
ff\delta $ interaction is the strongest (by symmetry). The symmetry  of
LaFe$_4$Sb$_{12}$ is I3\={m} and does not have the full set of 48 cubic
operations, but the effective mass of this specific band is very
isotropic, and rather small, $m^{*}\approx 0.2m_e.$ 

Let us now consider the second highest occupied band at the ${\bf \Gamma }$
point, which is a Fe/Co $t_{2g}$ band. With maximal cubic symmetry this band
would have been triply degenerate at ${\bf \Gamma }.$ Since the actual
symmetry of LaFe$_4$Sb$_{12}$ is lower, the corresponding state is split
into double degenerate and nongenerate states, the latter being the most
symmetric combination of the three $t_{2g}$ orbitals, $xy+yz+zx.$ This state
can also be written as $3z^{\prime 2}-r^2,$ where $z^{\prime }$ is the
threefold axis. $z^{\prime }$ is one of the three principal axis for the
mass tensor at the ${\bf \Gamma }$ point. Note that this state is directed
towards La and thus has little overlap with the $d$ orbitals on the
neighboring Fe/Co sites. This results in a low dispersion and high effective
mass\cite{mass}, $\langle m^{*}\rangle \approx 3.$
The top of this band is only 0.1 eV below the top of the previously
described Sb band. The combination of the two bands of so different
effective masses in direct vicinity of the Fermi level is what
provides high thermopowers in this material.

\begin{figure}[tbp]
\centerline{\epsfig{file=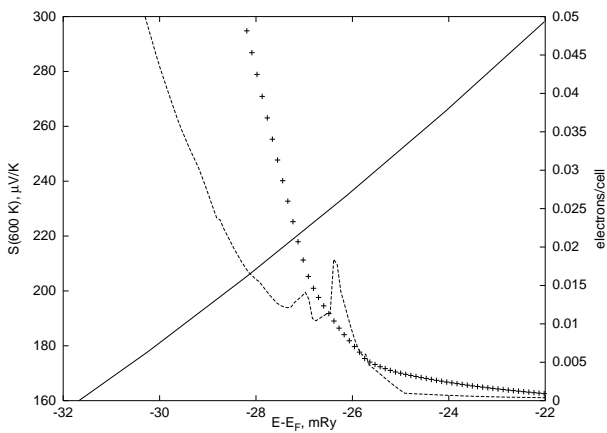,width=0.95\linewidth}}
\vspace{0.1in} \setlength{\columnwidth}{3.2in} \nopagebreak
\caption{Hole concentration (+), Hall number (dashed) and thermopower
at 600 K (solid) as a function of the Fermi level shift with
respect to the center of the gap (Fermi level for stoichiometric
La(Fe$_{0.75}$Co$_{0.25})_3$Sb$_{12}$.}
\label{tr} \end{figure}

In order to understand the transport more quantitatively, we have performed
calculations of transport properties\cite{calc} as a function of hole doping
away from the gap based on the 25\% Co virtual crystal band structure of
Fig. \ref{bands}. The Hall number, carrier concentration (doping level), and
thermopower at 600K are shown as a function of $E_F$ in Fig. \ref{tr}. Over
almost all of the range shown the Hall number is lower than the doping
level. This deviation, which increases with band filling, reflects
deviations from isotropic
parabolic band behavior. Although the presence of the heavy
band is responsible for the high thermopower of this material, $S$ at 600K
does not show any noticeable structure near $-26$ mRy, which is the onset of
the heavy band. This is because the form of Eq.(\ref{S}) has strong
contributions at energies up to 2-3 $kT$ from the chemical potential \cite
{22} and because $kT$ at 600 K is 3.7 mRy. Thus the heavy band contributes
to $S$ over the entire range of Fig.\ref{tr}. However, the carrier
concentration increases sharply below the onset of the second band,
reflecting its flat dispersion. This provides a mechanism for pinning the
Fermi level near the onset, providing high thermopowers even though the
carrier concentration is apparently difficult to control in these materials.

Although the Hall number is not reported, the sample shown by Sales, Mandrus
and Williams \cite{3} has $S$(600K)$\approx $180 V/K. Comparing with our
calculations, this corresponds to a chemical potential of $-30$ mRy, or 4
mRy below the onset of the heavy band. This implies a doping level of 0.08
holes per cell and a Hall number of 0.045 holes per cell. The corresponding
Hall concentration of $1.2\times 10^{20}$ holes/cm$^{-3}$ is consistent with
the reported lower end of the range of Hall concentrations reported in Ref. 
\onlinecite{3} for various samples ($1\times 10^{20}$ cm-3). 

\begin{figure}[tbp]
\centerline{\epsfig{file=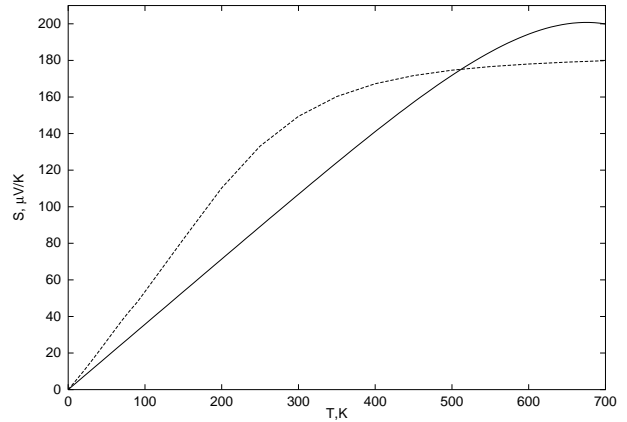,width=0.95\linewidth}}
\vspace{0.1in} \setlength{\columnwidth}{3.2in} \nopagebreak
\caption{ Calculated  at $E-E_F=30.3$ mRy 
(dashed) and experimental (solid) thermopower.}
\label{ST} \end{figure}

The temperature dependence of $S$ at a chemical potential $-$30.3 mRy
(matching the experimental value of $S$ at 500K) is shown in Fig. \ref{ST}.
The agreement is fair; the slope at low temperatures is
too high and the saturation temperature is too low, although the qualitative 
fact that $S$ is rather high even at high doping levels is reproduced.
One possible reason for
the quantitative disagreement is  neglect of the energy dependence of $\tau$.
 Since the two relevant
bands have very different characters (one is Sb $p$ and another is Fe/Co $d$%
), even
if the relaxation rates in each band are energy independent, as is likely,
they  may be different between the two bands, yielding an effective energy
dependence of $\tau$. This can change noticeably the
temperature dependence. Roughly speaking, the conductivity is defined mostly
by the light band, thus the thermopower is more or less additive for the two
bands. Because of very different dispersion laws the temperature dependence
of partial thermopower in each band is quite different, thus changing the
balance between the two can also change the net temperature dependence for
the total thermopower. There is, however, another, perhaps more important
effect, namely that the calculated band structure is distorted
relative to the samples due to the assumption of full La filling. 
We note the fact that the actual, as measured,
material is La deficient, and we observe from our calculations that  La $%
f_{xyz}$ orbitals are directly involved in the formation of the highest
occupied (``light'') band. 
The relative positions of the heavy and light bands, which are 
the key ingredient in determining the temperature dependence of $S$,
thus depend strongly on La filling.

Related to this it should be noted that the thermal conductivity for this
sample in the temperature range above 500 K is roughly $1/3$ electronic and $%
2/3$ lattice in origin. As mentioned $\kappa _l$ is strongly reduced from $%
\kappa _l$ of CoSb$_3$ by La addition, but most of this effect is expected
to occur at fairly low La concentrations within the rattling ion framework
of Slack and Cahill, Watson and Pohl \cite{15}, and supported by
measurements in the Ce(Fe,Co)$_4$Sb$_{12}$ on samples with different Ce
concentrations \cite{4}. Because of the strong interaction between the
valence bands and La, La vacancies should strongly scatter electrons
reducing the electrical conductivity. This is consistent with the wide range
in hole mobilities (2--30 cm$^2$/V$\cdot $s) measured for the various
samples. Based on this we conjecture that samples with higher La filling, if
they can be made, would have higher mobilities and higher values of $\sigma $
at a fixed band filling. Because of the Wiedemann-Franz relation, this would
also lead to higher $\kappa _e$. In that case, the value of $S$ would become
the most important factor determining $ZT$ at a given temperature. We
speculate it may be quite feasible to increase $ZT$ significantly in this
material if the La concentration could be made stoichiometric near 25\% Co
concentration. In this case, the doping level for maximum $ZT$ will occur at
lower carrier concentrations than any of the reported samples.
Depending on how much the mobility can be improved by La filling, it
may be possible to obtain high values $ZT$ at lower temperatures by doping
so that the chemical potential is roughly $kT$ above the onset of the heavy
band (i.e. in the 10$^{19}$ cm$^{-3}$ range).

To summarize, we have reported first principles calculations of band
structure and transport properties of La(Fe,Co)$_4$Sb$_{12}$. Despite high $%
ZT$ and large thermopower this material may be reasonably described as
metallic, both from the point of view of experimental measurements and the
calculated band structure. The electronic structure of La(Fe,Co)$_4$Sb$_{12}$
is strongly distorted from that of the corresponding binary, CoSb$_3$
implying an important  role of La in formation of the valence bands. This
fact finds its explanation in the specific character of the Sb states that form
the highest
valence band. Our results suggest that samples with higher La filling,
and lower hole concentrations may have even better TE properties.

This work was supported by the ONR and DARPA. Computations were performed
using the DoD HPCMO facilities at the NAVO and ASC. We are grateful for
helpful discussions with
T. Caillat,
J.L. Feldman,
J.P.  Fleurial,
D.T. Morelli,
L. Nordstrom,
W.E. Pickett,
G.A. Slack, and
T. Tritt.

\end{multicols}


\begin{references}
\bibitem{1}  G. Mahan, B. Sales and J. Sharp, Physics Today {\bf 50}, \#3, 42
(1997).

\bibitem{2}  T. Caillat, J.P. Fleurial and A. Borshchevsky in Proceedings
ICT'96 edited by T. Caillat, A. Borshchevsky and J.P. Fleurial (IEEE Press,
Piscataway, 1996), p. 151.

\bibitem{4}  J.P. Fleurial, A. Borshchevsky, T. Caillat, D.T. Morelli and
G.P. Meisner, {\it ibid}, p. 91.

\bibitem{3}  B.C. Sales, D. Mandrus and R.K. Williams, Science {\bf 272}, 1325
(1996).

\bibitem{5}  D.T. Morelli and G.P. Meisner, J. Appl. Phys. {\bf 77},
3777 (1995).

\bibitem{6}  L. Nordstrom and D.J. Singh, Phys. Rev. {\bf B 53}, 1103 (1996).

\bibitem{7}  D.J. Singh and W.E. Pickett, Phys. Rev. {\bf B 50}, 11235 (1994).

\bibitem{8}  J.L. Feldman and D.J. Singh, Phys. Rev. {\bf B 53}, 6273 (1996);
{\bf 54}, 712 (1996).

\bibitem{9}  D.J. Singh, Planewaves, Pseudopotentials and the LAPW Method,
Kluwer, Boston, 1994.

\bibitem{10}  J.M. Ziman, Principles of the Theory of Solids, Cambridge
University Press, Cambridge, 1972.

\bibitem{11}  C.M. Hurd, The Hall effect in Metals and Alloys, Plenum, New
York, 1972.


\bibitem{13}  A.E. Karakozov, I.I. Mazin and Y.A. Uspenski, Sov. Phys.
Doklady {\bf 277}, 848 (1984).

\bibitem{18}  D.T. Morelli, T. Caillat, J.P. Fleurial, A. Borshchevsky, J.
Vandersande, B. Chen and C. Uher, Phys. Rev. {\bf B 51}, 9622 (1995).

\bibitem{19}  D. Mandrus, A. Migliori, T.W. Darling, M.F. Hundley, E.J.
Peterson and J.D. Thompson, Phys. Rev. {\bf B 52}, 4926 (1995).

\bibitem{20}  T. Caillat, A. Borshchevsky and J.P. Fleurial, J. Appl. Phys.
{\bf 80}, 4442 (1996).

\bibitem{mass} For this band the mass tensor is relatively anisotropic.
Its principal values are $m_1^*=m_{z^{\prime }}^{*}\approx 3.9$, 
$m_2^*=m_3^{*}\approx 2.4.$
\bibitem{calc}  For the Hall coefficient we used Eq. 2.54 of
Ref.\onlinecite{11}.
777 first principle {\bf k}-point were used for
integration over the Brillouin zone both for the Hall coefficient and for
thermopower calculations.

\bibitem{22}  In metals convergence is usually obtained with an energy range
of 5 $kT$, but because of the strong variation of due to the band onsets we
used 10 $kT$ in the present calculations.

\bibitem{15}  D.G. Cahill, S.K. Watson and R.O. Pohl, Phys. Rev. {\bf 
B 46}, 6131
(1992).

\end{references}
\end{document}